\documentstyle[12pt,psfig]{article}
\newcommand{\simlt}{\,\,\raise2pt\hbox{$<$}\llap{\lower3pt\hbox{$\sim$}}\,\,}
\newcommand{\simgt}{\,\,\raise2pt\hbox{$>$}\llap{\lower3pt\hbox{$\sim$}}\,\,}
\long\def\comment#1{ }    
\def\alarm#1#2{#1_{\lower2pt\hbox{$\scriptstyle {\rm #2}$}}}
\def\ala#1#2{#1_{\lower2pt\hbox{$\scriptstyle {#2}$}}}
\def\alasrm#1#2{#1_{\lower1.5pt\hbox{$\scriptscriptstyle {\rm #2}$}}}
\def\alas#1#2{#1_{\lower1.5pt\hbox{$\scriptscriptstyle {#2}$}}}

\def\muR{\ala{\mu}{R}}
\def\muF{\ala{\mu}{F}}

\newcommand{\yourtitle}[1]{
\mbox{}\\
\vskip 4\baselineskip
{\bf\noindent #1}\\ }
\newcommand{\youraddress}[1]{
\noindent\mbox{}\hspace{1in}\parbox[t]{4.0in}{#1}\\ }
\newcommand{\yournames}[1]{
\mbox{}\\
\mbox{}\\
\vskip-0.8cm
\noindent\mbox{}\hspace{1in}{#1}\\ }
\newcommand{\yourabstract}[1]{
\mbox{}\\
\mbox{}\\
{\bf\noindent Abstract}\\
\begin{center}
\mbox{}\parbox[t]{5.in}{#1}
\end{center} }
\newcommand{\yoursection}[1]{
\vskip 2\baselineskip
{\bf\noindent #1}\\
\mbox{}\\
\vspace{-0.19in}}

\begin{document}
\vskip-2cm
\vskip1.2cm
\null\vskip-1cm

\yourtitle{EFFECTS OF SHADOWING ON DRELL--YAN DILEPTON PRODUCTION IN HIGH
ENERGY NUCLEAR COLLISIONS }
\yournames{K.J.~Eskola, V.J.~Kolhinen, and P.V.~Ruuskanen}
\youraddress{Department of Physics, University of Jyv\"askyl\"a \\
P.O.Box 35, FIN-40351, Jyv\"askyl\"a, Finland}
\yournames{R.~L.~Thews}
\youraddress{Department of Physics, University of Arizona,\\
Tucson, Arizona 85721, USA}
\yourabstract{
We compute cross sections for the Drell--Yan process in nuclear collisions at
next-to-leading order (NLO) in $\alpha_s$.  The effects of shadowing on the
normalization and on the mass
and rapidity dependence of these cross sections are presented.
An estimate of
higher order corrections is obtained from next-to-next-to-leading order
(NNLO) calculation of the rapidity-integrated mass distribution.
Variations in these predictions resulting from
choices of parton distribution sets are discussed.
Numerical results for mass distributions at NLO are presented
for RHIC and LHC energies,
using appropriate rapidity intervals.  The shadowing factors
in the dilepton mass range $2 < M < 10$ GeV are predicted to
be substantial , typically 0.5 - 0.7 at LHC, 0.7 - 0.9 at RHIC, and
approximately independent of the choice of parton distribution sets
and the order of calculation.
}

\clearpage
\yoursection{Introduction}

In a previous study \cite{9502372} we provided a systematic survey of
theoretical predictions for the Drell--Yan process \cite{DY,Neerven} in
nucleon--nucleon collisions at energies relevant to ion--ion experiments at
RHIC and LHC.  In this study we extend our work to nuclear collisions at the
same energies.  A short theoretical discussion can be found in \cite{9502372}
and a more complete review of the perturbative QCD calculations in van
Neerven's article \cite{Neerven}.  As before, we explore the uncertainties in
the dependence of the production rate on the dilepton's mass $M$ and rapidity
$y$ due to the choice of parton distribution set.  In addition, we
investigate thoroughly the
effects of nuclear shadowing \cite{EsKoRuSa}
on the normalization and on the mass and rapidity dependence of these cross
sections.

Our predictions for ${\rm d}\sigma^{AA}/{\rm d}M{\rm d}y$ are based on a
perturbative analysis of the underlying partonic processes to order
$\alpha_{\rm s}$ \cite{DY1,Neerven1,Neerven2,Neerven3}. The parton cross
sections are folded with parton distributions which are extracted from data
on the basis of next-to-leading order analysis. Unfortunately, only leading
order determination of nuclear shadowing effects is available presently,
partly due to the limited amount of data available for such a study.
Even though it deserves a careful study, we feel that the main
features of shadowing will not change qualitatively when going from leading
order to next-to-leading order analysis.

\begin{figure}[h]
\centerline{\psfig{figure=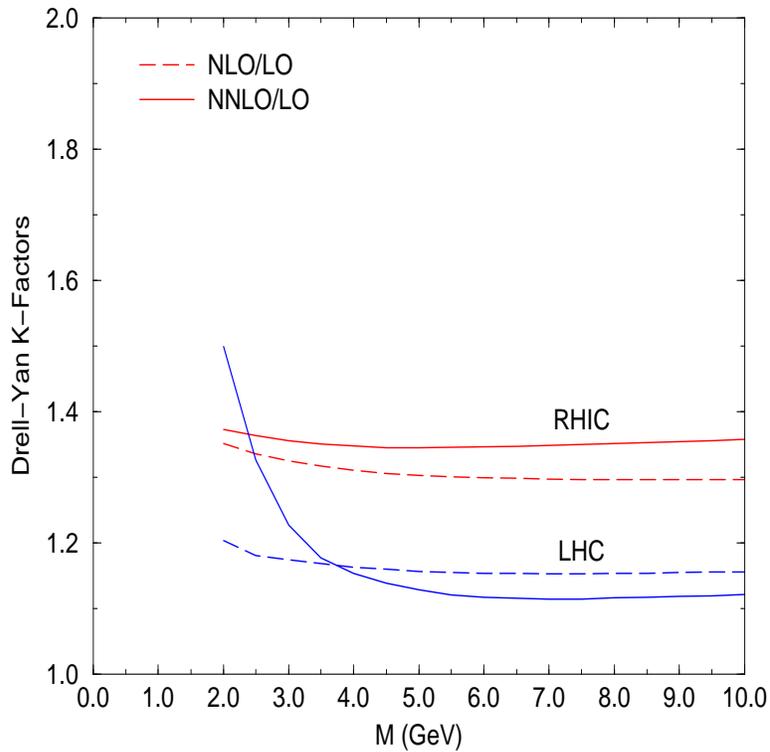,height=4in,width=4in}}
\vspace{0.0in}
\caption{$K$-factors at RHIC energy (200 GeV) and LHC energy (5.5 TeV)
for NLO (dashed lines) and NNLO (solid lines) Drell--Yan calculations
of ${\rm d}\sigma/{\rm d}M$ using MRSG structure functions in proton-proton
collisions.}
\label{figkfactor}
\end{figure}

As in earlier studies, we find that the perturbative corrections and the
uncertainties due to the choice of parton set grow as $M$ decreases.
From the point of view of the heavy ion physics, the mass region from 2
to 10 GeV is of most interest.  In order to estimate the applicability
of perturbative calculations at these relatively low mass values we
recall results for the rapidity-integrated ${\rm d}\sigma/{\rm d}M$, for
which the complete NNLO calculations (order $\alpha_{\rm s}^2$) are
available.  The relative magnitude of the NNLO correction sets one limit
on our confidence in the applicability of perturbation theory.  We show
in Figure \ref{figkfactor} the mass dependence of the $K$-factors from a
calculation with NLO parton distributions and parton level cross
sections up to NNLO.  One sees that for $M > 2$ GeV at RHIC energy and
for $M > 3.5$ GeV at LHC energy the NNLO contribution is at most about a
5 \% correction to the NLO result.  For lower masses at LHC energy the
NNLO contribution grows sharply.  At $M = 2$ GeV, the NNLO correction
alone is about 25 \%, which is larger than the NLO correction itself of
about 20 \%.  Hence in this mass and energy range we must consider the
calculations as merely a general indication of magnitudes, since
convergence of the QCD perturbative series is not yet evident.  Keeping
this caveat in mind, we restrict ourselves in the following calculations
to the NLO results.

We will not repeat here the discussion in \cite{9502372} on the
dependence of the cross sections on the renormalization scale $\muR$
and the factorization scale $\muF$, or choice of $\overline{\rm MS}$ or DIS
regularization scheme.  Since the determination of parton
distributions has improved considerably during last few years, we
calculate the NLO cross sections utilizing the recently-extracted
CTEQ \cite{CTEQ}, GRV \cite{GRV}, and MRS \cite{MRS} sets.  In this
connection we discuss also the
dependence of the expected shadowing effects on the choice of
parton distribution set.

In the next section we will define the average nucleon-nucleon cross section
in a nucleus-nucleus collision  and, using the nuclear transverse overlap
function, the number distributions normalized to one collision. We will also
briefly introduce the nuclear shadowing modifications to the parton
distributions.
In the following
section we present the NLO results at RHIC and LHC energies.  The effects
of shadowing are then implemented, followed by numerical results on the
rapidity-integrated cross sections.
A discussion
of general properties of the shadowing correction completes this work.

\yoursection{Cross sections and shadowing modifications}

Let us consider a collision of two nuclei with mass numbers $A_1$ and $A_2$.
Since the Drell--Yan process is sensitive to the isospin structure of the
nuclei, we must specify also the proton numbers $Z_1$ and $Z_2$,
and neutron numbers $N_1$ and $N_2$. We will
assume that the shapes of the proton and neutron density distributions within
a nucleus are the same:
\begin{eqnarray}
\alarm{\rho}{p}({\bf r}) &=& \alarm{\rho}{p}(z,{\bf s}) = \frac{Z}{A}
\ala{\rho}A({\bf r}), \nonumber \\
\alarm{\rho}{n}({\bf r}) &=& \alarm{\rho}{n}(z,{\bf s}) = \frac{N}{A}
\ala{\rho}A({\bf r}),
\end{eqnarray}
where $\ala{\rho}A({\bf r})$ is the total nucleon number density. In the
numerical calculations we use the Wood-Saxon parametrization
\begin{equation}
\ala{\rho}A({\bf r}) =\frac{n_0}{1+e^{(r-\ala{R}{A})/d}}
\label{eq:WS}
\end{equation}
where $n_0=0.17\ {\rm fm}^{-3},\ d=0.54\ {\rm fm}$ and $\ala RA =
1.12A^{1/3}-0.86A^{-1/3}\ {\rm fm}$.
We define the nuclear thickness function $\ala{T}{A_i}({\bf s})$
and the overlap function $\ala{T}{A_1A_2}({\bf b})$
at impact parameter ${\bf b}$ as
\begin{eqnarray}
\ala{T}{A}({\bf s})&=&\int {\rm d}z
\ala{\rho}{A}(z,{\bf s}), \nonumber \\
\ala{T}{A_1A_2}({\bf b})&=&\int {\rm d}{\bf s}
\ala{T}{A_1}({\bf s})\ala{T}{A_2}({\bf b-s})\,.
\label{eq:overlap}
\end{eqnarray}
In a $p-A$ collision we expect
the total cross section of a hard process to be the sum of individual
nucleon-nucleon cross sections,
\begin{equation}
{\rm d}\sigma_{{\rm p}A} =
Z{\rm d}\sigma_{\rm pp}+N{\rm d}\sigma_{\rm pn}.
\end{equation}
The corresponding cross section
${\rm d}{\sigma}_{A_1A_2}$
for an $A_1+A_2$ collision can be written as
\begin{equation}
{\rm d}\sigma_{A_1A_2}=
{Z_1}{Z_2}\,{\rm d}\sigma_{\rm pp}+
{N_1}{Z_2}\,{\rm d}\sigma_{\rm np}+
{Z_1}{N_2}\,{\rm d}\sigma_{\rm pn}+
{N_1}{N_2}\,{\rm d}\sigma_{\rm nn}\, .
\label{eq:sigmaAA}
\end{equation}
The average number distribution in one collision at impact parameter
${\bf b}$ is obtained by multiplying the average nucleon-nucleon cross
section in $A_1+A_2$ collision, ${\rm d}{\sigma}_{A_1A_2}^{\rm NN}= {\rm
d}{\sigma}_{A_1A_2}/A_1A_2$,
with the overlap function $T_{A_1A_2}({\bf b})$.
In particular, for the mass distribution per unit rapidity in
central, zero impact parameter collision we have
\begin{equation}
\frac{{\rm d}\ala{N}{A_1A_2}}{{\rm d}y{\rm d}M} =
\frac{\ala{T}{A_1A_2}(0)}{A_1A_2}\,
\frac{{\rm d}{\sigma}_{A_1A_2}}{{\rm d}y{\rm d}M}
\label{eq:dydM}
\end{equation}

In the QCD-based parton model the Drell--Yan cross section for
nucleon($n_1$)-nucleon($n_2$) collision ($n_i={\rm p,n}$) can be
expressed as
\begin{equation}
{\rm d}{\sigma}_{\!n_1\!n_2} = \sum_{i,j}
f_i^{n_1}(x_1,\mu^2)\otimes f_j^{n_2}(x_2,\mu^2)\otimes
{\rm d}\ala{\hat\sigma}{ij}
\label{eq:sigman1n2}
\end{equation}
in terms of partonic cross sections (or coefficient functions)
${\rm d}\ala{\hat\sigma}{ij}$ for partons $i$ and $j$ and parton distribution
functions $f_i^{n_1}(x,\mu^2)$ of parton $i$ in nucleon $n_1$.
In numerical calculations it is practical to combine eqs. (\ref{eq:sigmaAA})
and  (\ref{eq:sigman1n2}) to join the parton distributions in proton and
neutron using appropriate weights to a parton distribution in an average
nucleon.

The distribution of a parton flavour $i$ in a bound proton of a nucleus
$A$ can be written as $f_{i/A}^p(x,Q^2)=R_i^A(x,Q^2)  f_i^p(x,Q^2)$, where
$f_i^p(x,Q^2)$ is the corresponding parton distribution of the free
proton, and $R_i^A(x,Q^2)$ defines the nuclear effects (shadowing) at each
$x$ and $Q^2$.  The parton distributions of bound neutrons can then be
obtained by approximating $f_{u(\bar u)/A}^n\approx f_{d(\bar d)/A}^p$
and $f_{d(\bar d)/A}^n\approx f_{u(\bar u)/A}^p$
(exact for isoscalar nuclei) and $f_{i/A}^n=f_{i/A}^p$ for the isospin
symmetric distributions.

In practice, we include the nuclear effects in the computation of the
Drell--Yan dilepton cross sections ${\rm d}{\sigma}_{A_1A_2}^{\rm NN}$
by using the
EKS98-parametrization of $R_i^A(x,Q^2)$ \cite{EKS}. EKS98 is based on the
DGLAP analysis of Ref. \cite{EKR} where the nuclear parton distributions
are assumed to be factorizable above $Q_0=1.5$ GeV.  The existence of
nuclear effects at this scale is taken as a given fact (i.e. the origin is
not discussed) and the absolute nuclear parton distributions are evolved
to the region $Q>Q_0$ by using the standard lowest order leading twist
DGLAP equations.  Constraints for $R_i^A(x,Q^2)$ are given by the data on
the ratios of structure functions $F_2^{A_1}/F_2^{A_2}$ from deep
inelastic lepton-nucleus scatterings, by the data on Drell--Yan dilepton
production in proton-nucleus collisions, and by conservation of baryon
number and momentum. More detailed discussion of the the analysis
\cite{EKR} is found in \cite{EsKoRuSa}.

In \cite{EKS} it was shown that the ratios $R_i^A=f_{i/A}^p/f_{i}^p$ are
within a few percent the same for different sets of LO parton
distributions of the free proton. However, as the DGLAP analysis
\cite{EKR} was only performed in the LO, we have a slight inconsistency in
the NLO computation of the cross sections here.  The
difference to the NLO DGLAP analysis of the nuclear parton
distributions is nevertheless expected to be small in the ratios
$R_i^A$.

\yoursection{Sensitivity to Parton Distribution Functions}

\begin{figure}[h]
\centerline{\psfig{figure=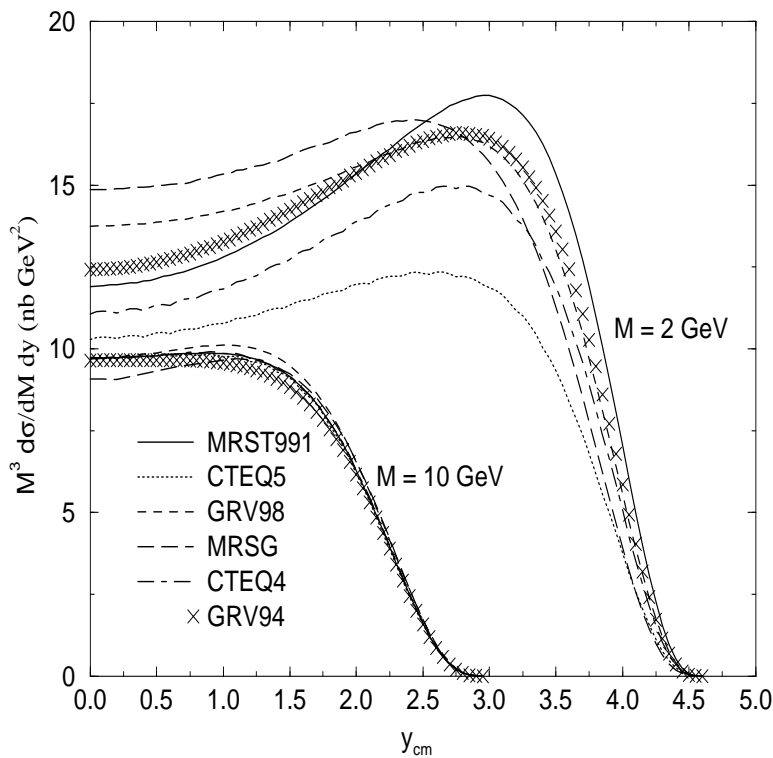,height=4in,width=4in}}
\vspace{0.0in}
\caption{NLO Drell--Yan rapidity distributions at $\sqrt s  = 200$ GeV in
p-p collisions for various parton distribution functions.}
\label{figrhicydist}
\end{figure}

To estimate the sensitivity of the Drell--Yan calculations to
the choice of parton distribution function sets, we have utilized the
most recent analyses of the three major groups (MRST99, CTEQ5, and
GRV98), where we have used the default parameters in the cases that
more than one option is available.  For further comparison, we have
also used the previous versions from these same groups (MRSG, CTEQ4,
and GRV94).

Shown in Figure \ref{figrhicydist} are the NLO differential
cross sections in center of mass rapidity for proton-proton collisions
at $\sqrt s = 200$ GeV at $M = 2$ and $M = 10$ GeV.
It is evident that at $M = 10$ GeV the $x$-values probed are
sufficiently large such that all distribution function sets are
constrained by data to very similar values.  (Recall that at LO the
partonic $x$-values $x_1$ and $x_2$ are fixed at
$x_{1,2} = {M \over \sqrt s} e^{\pm y}$.)
On the contrary, at
$M = 2$ GeV parton distributions at
the lower $x$-values are not sufficiently constrained
by data, and variations of 20 - 40 \% between sets are not uncommon.
Calculations with intermediate masses reveal that this large uncertainty
essentially disappears already when
$M = 4$ GeV, such that above this mass the variations are bounded
typically by 10 \%.

\begin{figure}[h]
\centerline{\psfig{figure=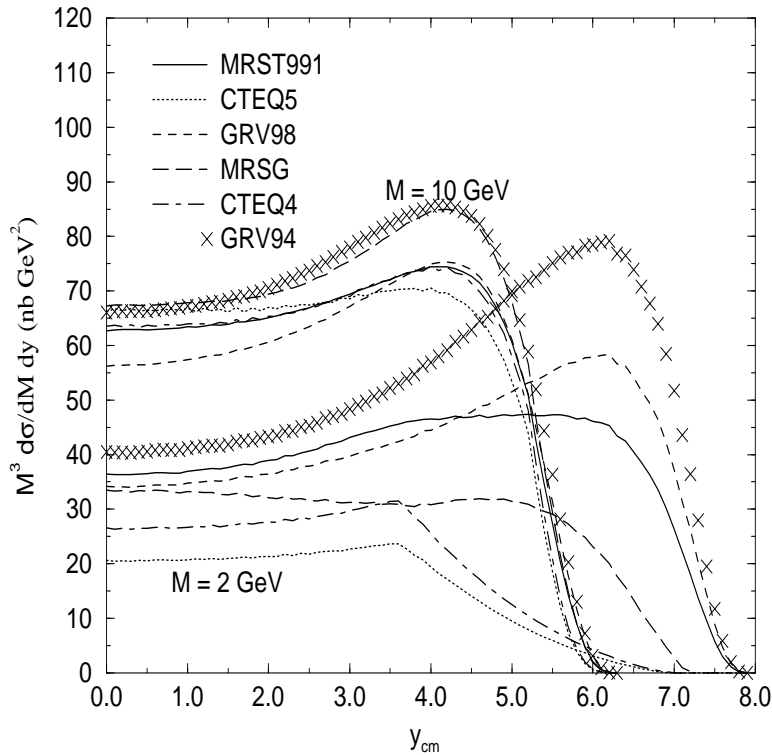,height=4in,width=4in}}
\vspace{0.0in}
\caption{NLO Drell--Yan rapidity distributions at $\sqrt s  = 5.5$ TeV in
p-p collisions for various parton distribution functions.}
\label{figlhcydist}
\end{figure}

Corresponding results for $\sqrt s = 5.5$ TeV are shown in
Figure \ref{figlhcydist}.
A similar pattern exists also at this energy, but with somewhat different
magnitudes.  At $M = 2$ GeV, the variation is at least a factor of 2.
(Note that the apparent discontinuity in slope for the CTEQ curves
are due to an absolute cutoff below a minimum $x$-value of $10^{-5}$.)
As one increases $M$ this uncertainly again decreases rapidly, reaching
the 10-15 \% level at $M = 10$ GeV.

\yoursection{Shadowing effects in rapidity distributions}

\begin{figure}[!p]
\centerline{\psfig{figure=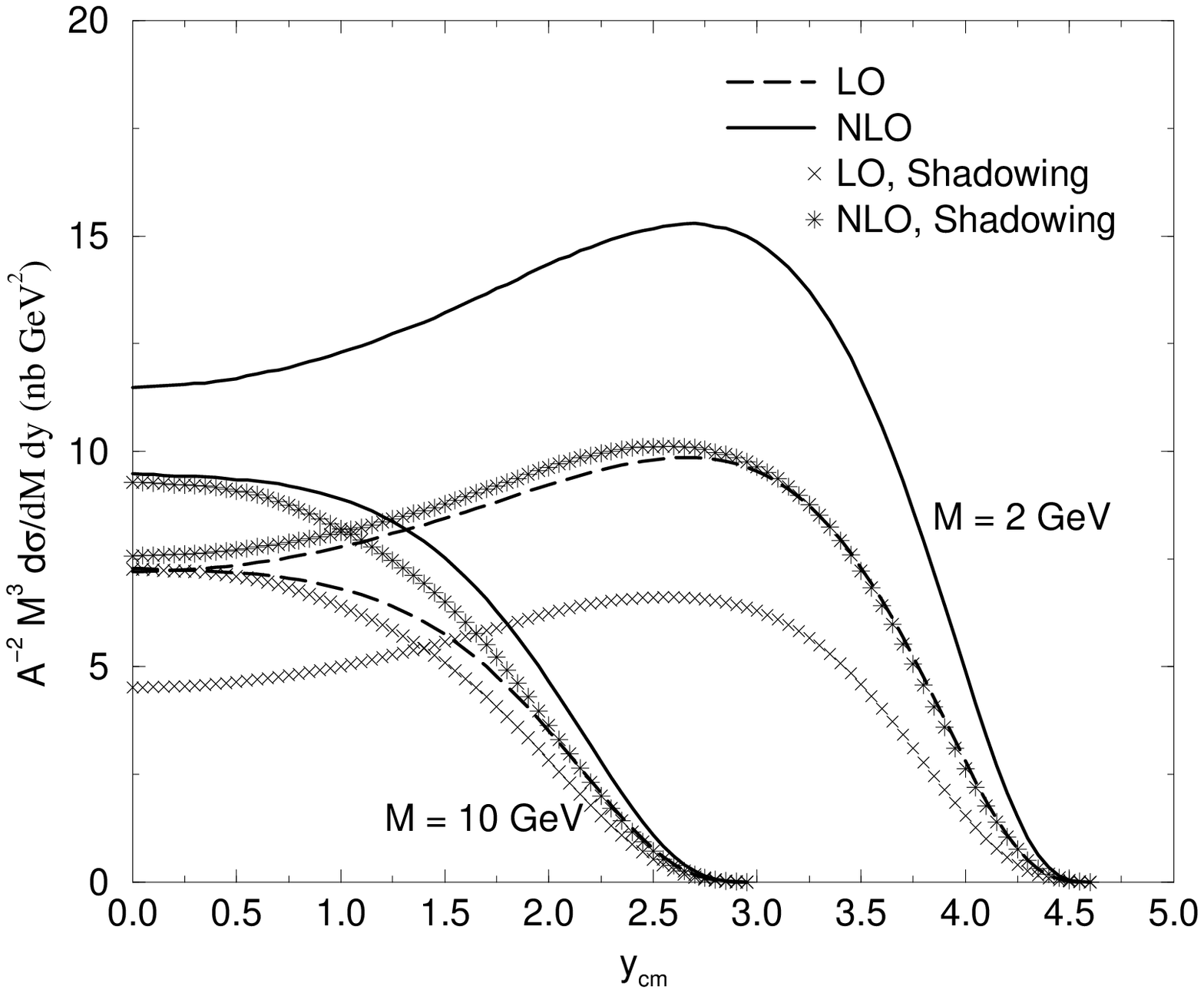,height=4in,width=4in}}
\vspace{0.0in}
\caption{Shadowing in LO and NLO Drell--Yan rapidity distributions
calculated with MRST99 structure functions for
Au-Au collisions at RHIC with $\sqrt s  = 200$A GeV.}
\label{figrhicshad}
\end{figure}

For the effects of shadowing at RHIC and LHC, we choose a representative
parton distribution set, the MRST99.
Shown in Figure \ref{figrhicshad} are Au-Au LO and NLO cross sections per
nucleon at RHIC for $M = 2$ GeV and $M = 10$ GeV, both with and
without shadowing.
One sees that at smaller $M$  the effects of shadowing are quite large,
and in much of the kinematic range the magnitude is similar to that of
the NLO corrections.
\begin{figure}[!p]
\centerline{\psfig{figure=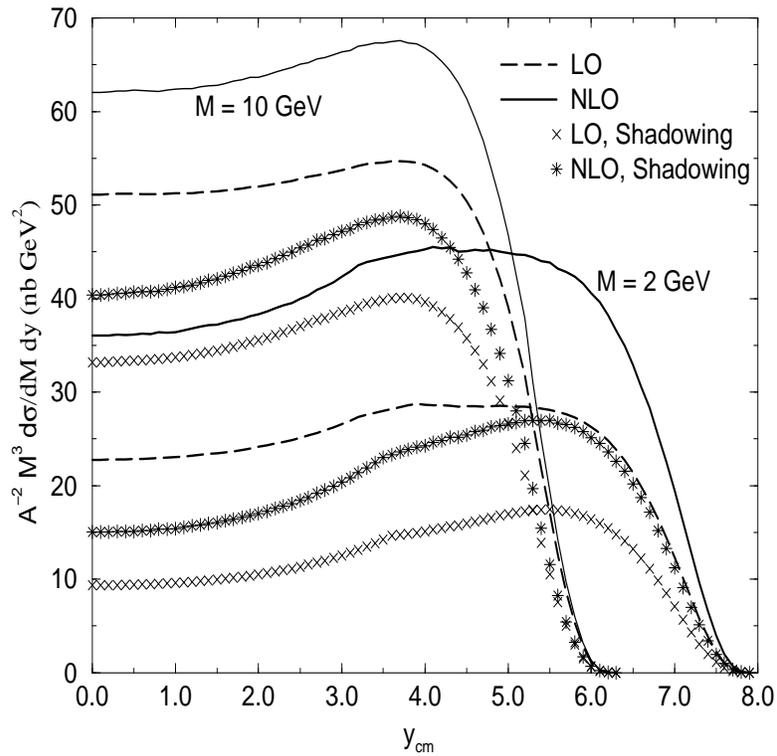,height=4in,width=4in}}
\vspace{0.0in}
\caption{Shadowing in LO and NLO Drell--Yan rapidity distributions
calculated with MRST99 structure functions for
Pb-Pb collisions at LHC with $\sqrt s  = 5.5$A TeV.}
\label{figlhcshad}
\end{figure}
\begin{figure}[!p]
\centerline{\psfig{figure=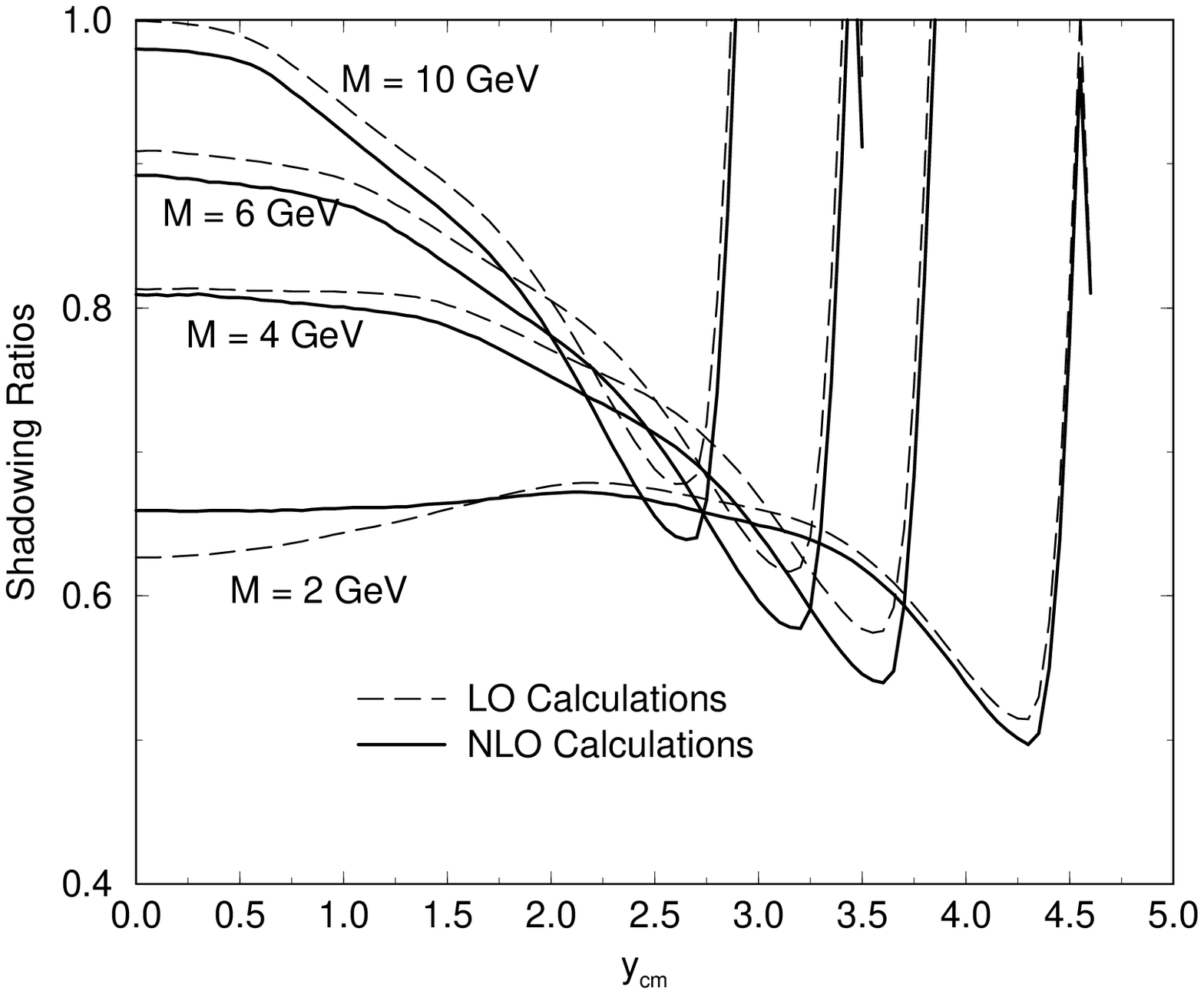,height=3.5in,width=3.5in}}
\vspace{0.0in}
\caption{Ratio of Shadowing/No Shadowing in LO
and NLO Drell--Yan cross sections ${\rm d}\sigma/{\rm d}M{\rm d}y$
as a function of rapidity for fixed values of $M$
at RHIC with $\sqrt s  = 200$A GeV.}
\label{figrhicshadratio}
\vskip 1.0truecm
\centerline{\psfig{figure=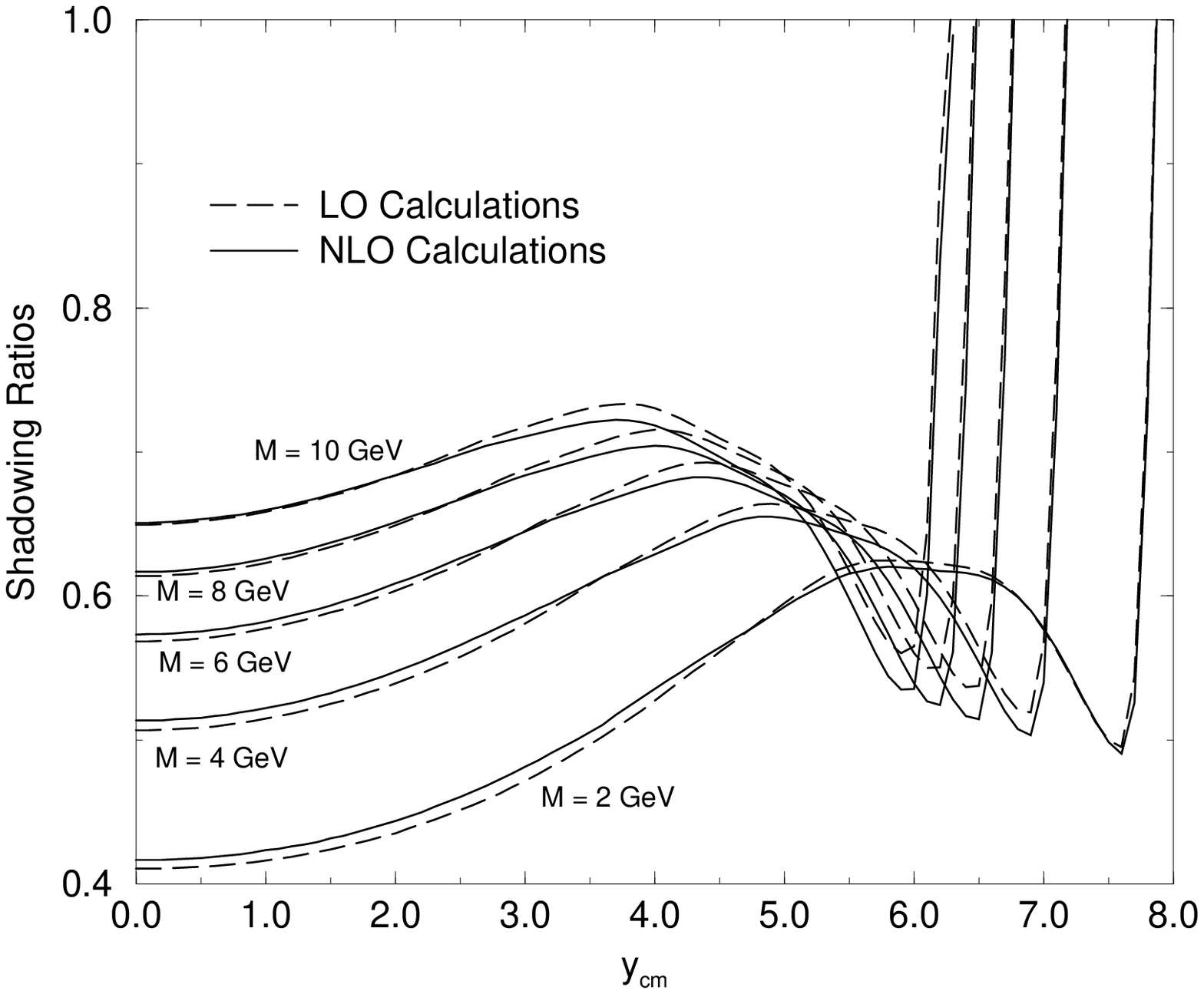,height=3.5in,width=3.5in}}
\vspace{0.0in}
\caption{Ratio of Shadowing/No Shadowing in LO
and NLO Drell--Yan cross sections ${\rm d}\sigma/{\rm d}M{\rm d}y$
as a function of rapidity for fixed values of $M$
at LHC with $\sqrt s  = 5.5$A TeV.}
\label{figlhcshadratio}
\end{figure}
\vfil\newpage

Figure \ref{figlhcshad}
presents the same calculations at LHC for Pb-Pb collisions.
The magnitudes of the shadowing corrections are even larger here,
due to the smaller target $x$-values probed at this energy
and rapidity range.

We also show in Figures \ref{figrhicshadratio} and \ref{figlhcshadratio}
the ratio of cross sections with and without shadowing corrections.
The dashed lines for LO are determined entirely by the fixed $x_{1,2}$
-values and the corresponding quark and antiquark shadowing ratios
$R_i^A(x,Q^2=M^2)$ for each nucleus.

The solid lines for the NLO ratios are not identical to those for LO, but
are seen to follow those curves quite closely.  This behavior
is somewhat unexpected,
since the NLO subprocesses
involve integrals over $x$-values significantly different than the
$x_{1,2}$.  It
cannot be explained by simple magnitude arguments, since the NLO
$K$-factors
are substantial, and also involve gluon structure functions not present
in the LO contribution.  What is evidently happening is that the NLO
integrals can be approximated by mean values of the integrand
at effective
$x$-values close to the $x_{1,2}$.  In any event, it appears that one
may be able to extract overall multiplicative shadowing factors which
will be approximately independent of the order of perturbative QCD and
the input structure functions.

\yoursection{Mass distributions}

\begin{figure}[!p]
\centerline{\psfig{figure=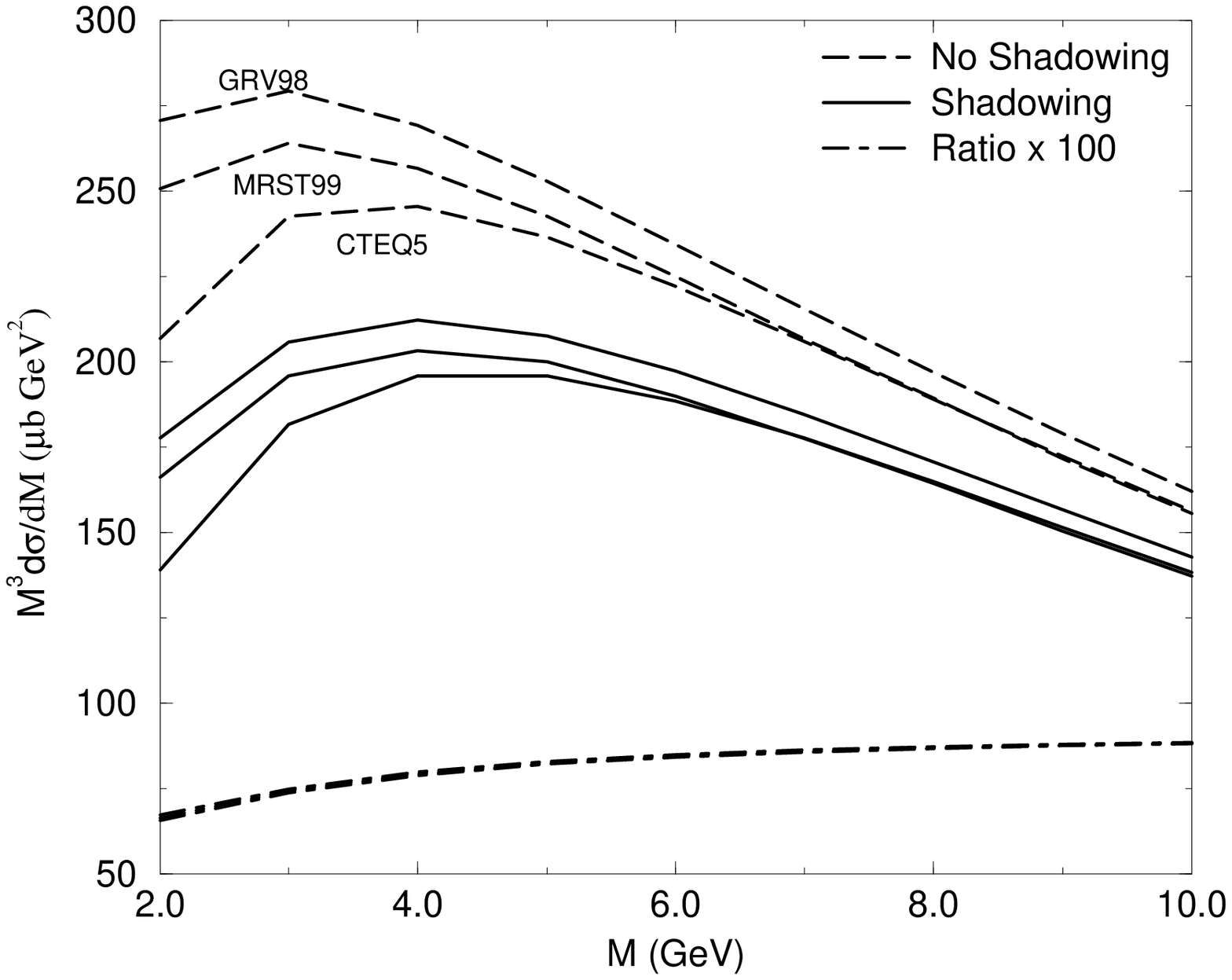,height=4in,width=4in}}
\vspace{0.0in}
\caption{Rapidity-Integrated $1.1 < y < 1.6$ NLO Drell--Yan Mass
distributions for Au-Au collisions at RHIC with $\sqrt s  = 200$A GeV.}
\label{figrhicdsdm}
\vskip 1.0truecm
\centerline{\psfig{figure=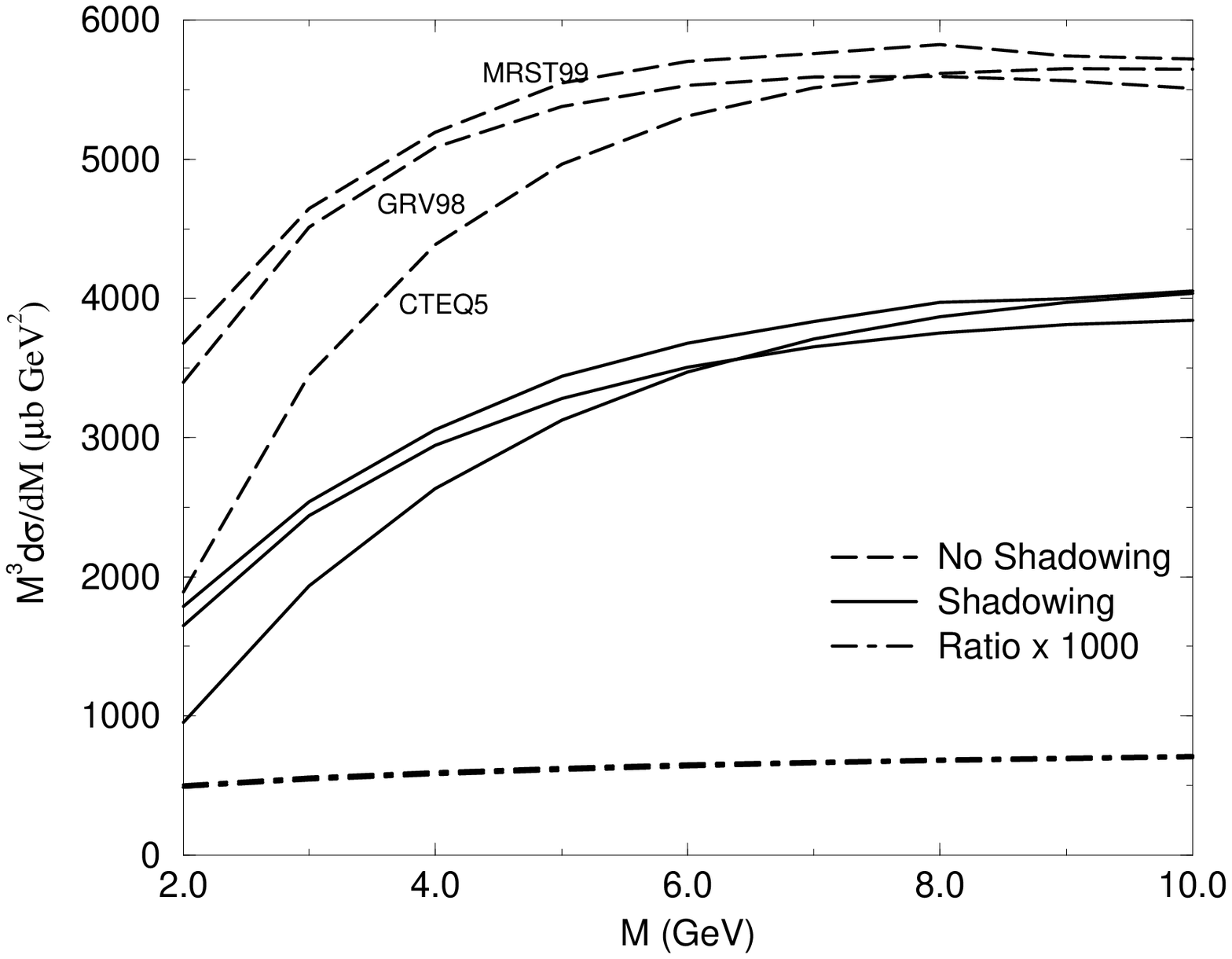,height=4in,width=4in}}
\vspace{0.0in}
\caption{Rapidity-Integrated $2.0 < y < 4.0$ NLO Drell--Yan Mass
distributions for Pb-Pb collisions at LHC with $\sqrt s  = 5.5$A TeV.}
\label{figlhcdsdm}
\end{figure}

We now calculate the expected mass distributions of the NLO nucleus-nucleus
cross sections
at RHIC and LHC, integrated over rapidity intervals appropriate
for acceptance of the detectors (1.1 - 1.6 at RHIC and
2.0 - 4.0 at LHC).  The results for RHIC are shown in
Figure \ref{figrhicdsdm} for the MRST99, CTEQ5, and GRV98
structure functions.  Both shadowing and no shadowing
calculations are presented.  Although there is a considerable
difference between results with different structure functions, especially
at low mass, this uncertainty is less than the expected differences
between shadowing and no shadowing. Also shown are the ratios of
shadowing to no shadowing predictions for calculations using
each of the three structure functions.  As anticipated, this
results in a universal shadowing curve, approximately independent
of structure function.  The shadowing curve at the bottom of the
figure actually displays all three results - the differences are
smaller than the width of the lines.

Corresponding results for LHC are shown in
Figure \ref{figlhcdsdm}.  The patterns are very similar.
At the lowest mass there is somewhat more dispersion due to
choice of structure function, but we already know that in this region
both the reliability of pQCD at NLO is quite weak and the parton
distributions are less constrained. The shadowing ratios converge again
to a universal curve.

For archival purposes, we present in Table \ref{tabsig} the calculated
NLO nucleus-nucleus cross section values for all three choices of
structure functions, both with and without shadowing.  In addition,
we list the corresponding p-Au (at RHIC) and p-Pb (at LHC) NLO
cross sections for the same detector rapidity interval in both the proton
and nucleus directions. (At ALICE we take into account the asymmetric
beam energies, and assume that the rapidity interval in the nucleus
direction is reached by interchanging the beam directions.)

\vfil\newpage
\yoursection{Conclusions}

We have studied the production of Drell--Yan dileptons at energies and
phase space regions appropriate for the BNL-RHIC and CERN-LHC heavy ion
experiments.  The NNLO calculations, available for ${\rm d}\sigma/{\rm
d}M$, indicate that a perturbative calculation converges well for
$M\simgt 4$ GeV, the NNLO correction being $\simlt 5$ \%.
At the LHC energy it becomes rapidly less reliable
as $M$ approaches 2 GeV.  Using a representative selection of most
recent parton distribution sets, MRS991 \cite{MRS}, CTEQ5 \cite{CTEQ},
and GRV98 \cite{GRV}, we found
that the differences between the sets lead to an uncertainty in the
cross sections which at small masses, $M\simeq 2$ GeV, is typically
20~--~40~\% at RHIC but around a factor of two at LHC.  This reflects the
lack of precision experimental constraints on parton distributions at small
$x$. At RHIC the results from different sets converge as $M$ increases,
but at LHC persist at the 10~--~15~\% level even for $M=10$ GeV.
We investigated the nuclear effects
using the parton distribution functions for free nuclei modified with
multiplicative shadowing functions determined from the available
experimental information \cite{EKS}. At RHIC energy, shadowing reduces
the cross section at $M\simeq 2$ GeV by a universal (i.e.
independent of structure function set or order or perturbative QCD)
factor of $\sim 35$ \%.  This factor
approaches $\sim 10$ \% for $M\simeq 10$ GeV. At LHC the reduction
for small masses is $\sim$ 55 \% and more than 30~\% even at $M=10$ GeV.
In summary, these calculations are limited by uncertainties due to
convergence of the perturbative series and parton distributions which
become large at small values of $M/\sqrt s$.
Thus they are generally not a problem for
all of the RHIC results, but become severe at LHC for $M\simlt 4$ GeV.
The overall effects of shadowing at RHIC energies
are smaller in magnitude than those at
LHC, but the change of the mass distribution shape
due to shadowing is a larger
effect at RHIC.

\vfil\newpage
\begin{table}[t]
\centerline{
\begin{tabular}{|c|c|c|c|c|c|c|} \hline \hline
\multicolumn{7}{|c|}{ Rapidity-integrated Drell--Yan cross section
{\hbox{$\displaystyle{M^3\frac{{\rm d}\sigma}
{{\rm d}M}}$} [$\mu$b\,GeV$^2$] }}  \\
\hline \hline
 & & & & & &  \\
 $M$ [GeV]& MRST99 & CTEQ5 & GRV98 & MRST99 & CTEQ5 & GRV98 \\
 & & & & Shadow& Shadow& Shadow   \\
 & & & & & &  \\
\hline
\hline
\multicolumn{7}{|c|}{$\sqrt s = 200$ GeV, $1.1 < y < 1.6$ for Au-Au at RHIC} \\
\hline
\hline
  2.0& 251& 207& 271& 166& 139& 178  \\ \hline
  3.0& 264& 243& 279& 196& 181& 206  \\ \hline
  4.0& 257& 246& 269& 203& 196& 212  \\ \hline
  5.0& 243& 236& 253& 200& 196& 208  \\ \hline
  6.0& 225& 222& 234& 190& 189& 197  \\ \hline
  7.0& 207& 206& 216& 177& 178& 185  \\ \hline
  8.0& 189& 189& 197& 164& 165& 171  \\ \hline
  9.0& 172& 172& 179& 150& 152& 157  \\ \hline
 10.0& 155& 156& 162& 137& 138& 143  \\ \hline
\hline
\hline
\multicolumn{7}{|c|}{$\sqrt s = 5.5$ TeV, $2.0 < y < 4.0$ for Pb-Pb at LHC} \\
\hline
\hline
  2.0& 3680& 1890& 3400& 1790&  954& 1650  \\ \hline
  3.0& 4640& 3450& 4510& 2540& 1930& 2440  \\ \hline
  4.0& 5190& 4390& 5080& 3060& 2630& 2940  \\ \hline
  5.0& 5550& 4960& 5380& 3440& 3120& 3280  \\ \hline
  6.0& 5700& 5310& 5530& 3680& 3470& 3500  \\ \hline
  7.0& 5760& 5510& 5590& 3830& 3700& 3650  \\ \hline
  8.0& 5820& 5620& 5590& 3970& 3870& 3750  \\ \hline
  9.0& 5740& 5650& 5560& 4000& 3970& 3810  \\ \hline
 10.0& 5720& 5640& 5510& 4050& 4030& 3840  \\ \hline
\hline
\hline
\multicolumn{7}{|c|}
{$\sqrt s = 200$ GeV, $1.1 < y < 1.6$ for p-Au, Au-p at RHIC} \\
\hline
\hline
  2.0& 1.32, 1.28& 1.10, 1.05& 1.44, 1.38& .937, 1.20& .781, .992& 1.01, 1.28  \\ \hline
  3.0& 1.42, 1.35& 1.31, 1.23& 1.50, 1.42& 1.06, 1.33& .987, 1.22& 1.12, 1.40  \\ \hline
  4.0& 1.40, 1.31& 1.34, 1.25& 1.47, 1.37& 1.09, 1.33& 1.05, 1.27& 1.14, 1.38  \\ \hline
  5.0& 1.34, 1.24& 1.31, 1.20& 1.40, 1.29& 1.08, 1.27& 1.06, 1.23& 1.12, 1.31  \\ \hline
  6.0& 1.27, 1.15& 1.25, 1.13& 1.32, 1.19& 1.04, 1.17& 1.03, 1.16& 1.08, 1.22  \\ \hline
  7.0& 1.19, 1.05& 1.18, 1.04& 1.23, 1.09& .995, 1.08& .991, 1.07& 1.03, 1.12  \\ \hline
  8.0& 1.11, .960& 1.10, .955& 1.14, .999& .942, .979& .938, .976& .976, 1.02  \\ \hline
  9.0& 1.02, .870& 1.02, .869& 1.06, .907& .881, .883& .881, .884& .914, .919  \\ \hline
 10.0& .942, .786& .940, .787& .974, .819& .822, .796& .821, .797& .850, .828  \\ \hline
\hline
\hline
\multicolumn{7}{|c|}
{$\sqrt s = 8.8$ TeV, $2.0 < y  < 4.0$ for p-Pb, Pb-p at LHC} \\
\hline
\hline
  2.0& 20.3, 20.1& 10.0, 8.24& 18.5, 19.6& 12.5, 15.8& 6.20, 6.75& 11.4, 15.5 \\ \hline
  3.0& 26.3, 26.6& 19.6, 17.3& 25.5, 26.8& 16.8, 22.7& 12.6, 15.1& 16.3, 22.7 \\ \hline
  4.0& 30.2, 30.9& 25.9, 23.9& 29.4, 30.8& 19.9, 27.6& 17.1, 21.7& 19.3, 27.2 \\ \hline
  5.0& 32.9, 33.9& 30.1, 28.7& 31.7, 33.0& 22.1, 31.2& 20.3, 26.8& 21.2, 30.0 \\ \hline
  6.0& 34.4, 35.4& 32.8, 31.9& 32.9, 34.2& 23.5, 33.4& 22.5, 30.4& 22.4, 31.8 \\ \hline
  7.0& 35.2, 36.1& 34.6, 33.7& 33.6, 34.9& 24.4, 34.7& 24.0, 32.7& 23.2, 33.0 \\ \hline
  8.0& 36.1, 36.8& 35.7, 34.7& 34.0, 35.1& 25.4, 35.8& 25.1, 34.0& 23.7, 33.7 \\ \hline
  9.0& 36.0, 36.4& 36.4, 35.1& 34.1, 35.1& 25.6, 35.8& 25.9, 34.8& 24.1, 34.1 \\ \hline
 10.0& 36.4, 36.4& 36.8, 35.3& 34.0, 35.0& 26.0, 36.2& 26.4, 35.3& 24.2, 34.2 \\ \hline
\hline
\end{tabular}}
\caption{NLO Drell--Yan Calculations for RHIC and LHC}
\label{tabsig}
\end{table}
\clearpage


\begin{thebibliography}{999}
%
\bibitem{9502372} S.~Gavin, S.~Gupta, R.~Kauffman, P.~V.~Ruuskanen,
D.~K.~Srivastava and R.~L.~Thews,
Int. J. Mod. Phys. {\bf A10} (1995) 2961.
%
\bibitem{DY} S.D.~Drell and T.M.~Yan, Phys.~Rev.~Lett.~{\bf 25} (1970)
316.
\bibitem{Neerven} W.L.~van~Neerven,
Int. J. Mod. Phys. {\bf A10} (1995) 2921.
%
\bibitem{EsKoRuSa}
        K.J. Eskola, H. Honkanen, V.J. Kolhinen, P.V. Ruuskanen and C.A.
Salgado, this volume.
%
\bibitem{DY1} 
K.~Kajantie, J.~Lindfors and R.~Raitio, Phys.~Lett.\ {\bf 74B} (1978) 384; \\
J.~Kubar-Andre and F.E.~Paige, Phys.~Rev.~{\bf D19} (1979) 221;\\
G.~Altarelli, R.K.~Ellis and G.~Martinelli, Nucl.~Phys.~{\bf B157} (1979)
461;\\
J.~Kubar, M.~le~Bellac, J.L.~Munier and G.~Plaut, Nucl.~Phys.~{\bf B175}\\
(1980) 251; B.~Humpert and W.L.~van~Neerven, Nucl.~Phys.~{\bf B184} (1981)
225.
%
 \bibitem{Neerven1}  
R.~Hamberg, W.L.~van~Neerven and T.~Matsuura, Nucl.~Phys.~{\bf B359} (1991)
343; \
W.L.~van~Neerven and E.B.~Zijlstra, Nucl. Phys. {\bf B382} (1992) 11.
%
\bibitem{Neerven2} 
T. Matsuura and W.L.~van~Neerven, Z.~Phys.~{\bf C38} (1988) 623;\\
T. Matsuura, S.C.~van~der~Marck and W.L.~van~Neerven,
Nucl.~Phys.~{\bf B319} (1989) 570; Phys.~Lett.~{\bf 211B} (1988) 171.
%
\bibitem{Neerven3} P.J.~Rijken and  W.L.~van~Neerven,
Phys.~Rev.~{\bf D51} (1995) 44.
%
\bibitem{CTEQ} H. L. Lai, J. Huston, S. Kuhlmann, J. Morfin,
F. Olness, J. F. Owens, J. Pumplin and W. K. Tung (CTEQ Collaboration),
Eur.~Phys.~J.~{\bf C12} (2000) 375-392 [hep-ph/9903282].
%
\bibitem{GRV} M.~Gl\" uck, E.~Reya and A.~Vogt,
Eur.~Phys.~J.~{\bf C5} (1998) 461-470
        [hep-ph/9806404].
%
\bibitem{MRS} A.D.~Martin, W.J.~Stirling, R.G.~Roberts and R.S.~Thorne,
Eur.~Phys.~J.~{\bf C4} (1998) 463;
[hep-ph/9907231].
%
\bibitem{EKS}
        K.J. Eskola, V.J. Kolhinen and C.A. Salgado,
        Eur. Phys. J. C9 (1999) 61-68 [hep-ph/9807297];
        http://www.phys.jyu.fi/research/urhic/eks98param.html;
        H. Plothow-Besch, {\it 'PDFLIB: Proton, Pion and Photon Parton Density
        Functions, Parton Density Functions of Nucleus, and $\alpha_s$
        Calculations'}, User's Manual -- Version 8.04, W5051 PDFLIB 2000.04.17
        CERN--ETT/TT.
%
\bibitem{EKR}
        K.J. Eskola, V.J. Kolhinen and P.V. Ruuskanen,
        Nucl. Phys. B535 (1998) 351-371 [hep-ph/9802350].
%


\end{thebibliography}
\end{document}